\def \be {\begin{equation}}
\def \eq {\end{equation}}
\def \bee {\begin{eqnarray}}
\def \eqq {\end{eqnarray}}
\def \nn {\nonumber}
\def \bea {\begin{array}{c}}
\def \eqa {\end{array}}

\def \la {\langle}
\def \ra {\rangle}

\def \C {{\bf C}}
\def \Z {{\bf Z}}

\def \del {\partial}
\def \dels {\partial\kern-.5em / \kern.5em}
\def \As {{A\kern-.5em / \kern.5em}}
\def \Ds {D\kern-.7em / \kern.5em}
\def \Psib {\bar{\Psi}}

\def \a {\alpha}
\def \b {\beta}
\def \g {\gamma}
\def \G {\Gamma}
\def \d {\delta}
\def \eps {\epsilon}
\def \m {\mu}
\def \n {\nu}

\def \lam {\lambda}

\def \s {\sigma}
\def \r {\rho}

\def \O {\Omega}
\def \th {\theta}
\def \Th {\Theta}

\def \cD {{\cal D}}

\def \Psib {\bar{\Psi}}

\def \epsb {\bar{\epsilon}}
\def \II {I\hspace{-.1em}I\hspace{.1em}}
\def \IIA {\mbox{\II A\hspace{.2em}}}

\def \yd {y^{\dagger}}
\def \thd {\th^{\dagger}}

\def \rd {\dot{\r}}
\def \sd {\dot{\s}}
\def \one {{\bf 1}}

\def \cDb {\bar{\cD}}
\def \yb {\bar{y}}
\def \zb {\bar{z}}
\def \phih {\hat{\phi}}
\def \ve {\varepsilon}

\documentstyle[12pt,amssymbols]{article}

\begin{document}
\begin{titlepage}
\today          \hfill
\begin{center}
\hfill    UU-HEP/97-02\\
\hfill    EFI-97-24\\

\vskip .5in

{\large $p$-$p'$ Strings in M(atrix) Theory }

\vskip .5in
Pei-Ming Ho$^1$, Miao Li$^2$ and Yong-Shi Wu$^1$ \\
\vskip .2in
{\em $^1$ Department of Physics,
University of Utah \\
Salt Lake City, Utah 84112}
\vskip .2in
{\em $^2$ Enrico Fermi Institute and Department of  Physics \\
University of Chicago,
5640 S Ellis Ave., Chicago, IL 60637}

\end{center}

\vskip .5in

\begin{abstract}

We study the off-diagonal blocks in the M(atrix)
model that are supposed to correspond to open
strings stretched between a D$p$-brane and a
D$p'$-brane. It is shown that the spectrum,
including the quantum numbers, of the
zero modes in the off-diagonal blocks
can be determined from the index theorem
and unbroken supersymmetry, and indeed
reproduces string theory predictions for
$p$-$p'$ strings. Previously the matrix
description of a longitudinal fivebrane
needed to introduce extra degrees of freedom
corresponding to 0-4 strings by hand. We show
that they are naturally associated with the
off-diagonal zero modes, and the supersymmetry
transformation laws and low energy effective
action postulated for them are now derivable
from the M(atrix) theory.

\end{abstract}
\end{titlepage}

\newpage
\renewcommand{\thepage}{\arabic{page}}
\setcounter{page}{1}
%THIS IS PAGE 1 (INSERT TEXT OF REPORT HERE)

\section{Introduction}

D(irichlet)-branes have many faces. In string theory,
they arise as nonperturbative dynamic objects, allowing
strings to end on them and carrying  R(amond)-R(amond)
charge \cite{Pol}. In the conformal field theory
formulation, a D$p$-brane is a $p$-dimensional
hyperplane in target space on which strings
satisfy the Dirichlet boundary conditions
\cite{DLP}. In the low energy field theory limit
(supergravity), it appears as a soliton-like background
with nontrivial R-R antisymmetric tensor field, solving
classical equations of motion (see a recent review
\cite{LP} and references therein). The low energy
dynamics of  parallel D-branes, due to strings stretched
between them, can be described by a dimensionally
reduced supersymmetric Yang-Mills theory on their
world volume \cite{Wit}, which happens to
describe a quantum space in the sense of
non-commutative geometry \cite{HW}.

In the M(atrix) model \cite{BFSS} for M theory,
which is conjectured to unify all known perturbative
string theories, the D0-branes are treated as
fundamental microscopic degrees of freedom. The
SYM quantum mechanics, which was originally
thought to be the low-energy theory of $N$
D0-branes, is promoted in the large $N$ limit
to the status of the fundamental light-cone
dynamics of M theory. As dimensionally reduced
$U(N)$ SYM theory, its field content matches
the lowest modes of open strings ending on D0-branes.
Thus, in M(atrix) theory, everything else appears
as a collective (bound) state of D0-branes. In
particular, a multiple parallel D-brane background is
realized as a block-diagonal matrix \cite{BFSS},
each block represented by a topologically nontrivial
gauge field configuration \cite{BSS,GRT} on a
D-brane volume. In this paper we study the dynamics
of D-branes by introducing and examining off-diagonal
blocks, that are supposed to correspond to strings
stretched between D-branes. One of the advantages
of the M(atrix) theory is that it provides a unifying
framework for explicitly dealing with both D-brane
backgrounds and strings stretched between them.

Previously Berkooz and Douglas \cite{BD} have
considered the background of a longitudinal M5-brane,
which wraps around the (invisible) 11-th direction that
defines the light-cone to give rise to a D4-brane
in \IIA language. They bypassed the question of
explicitly representing the D4-brane in matrix form,
but rather proposed a modified M(atrix) theory by
introducing by hand additional dynamical variables
that are supposed to correspond to the massless
modes of open strings stretched between the D4-brane
and the D0-branes (called 0-4 strings).
It was shown that integrating out the extra variables
leads to the correct gravitational field of an
M5-brane. Later Dijkgraaf, Verlinde and
Verlinde \cite{DVVa} showed that if one
integrates out the off-diagonal blocks in the
$U(N)$ matrix fields with two diagonal
blocks for a D4-brane and a D0-brane
respectively, one can also recover the
gravitational field of a longitudinal M5-brane.
Based on this result, one may be tempted to
identify the extra fields introduced in Ref.
\cite{BD} with the above-mentioned off-diagonal
blocks. However, there is a mismatch for the
quantum numbers: the extra bosonic field in
Ref. \cite{BD} is a spinor of the $SO(4)$
in the 4-brane directions, in accordance with
string theory \cite{Pol1}, while the bosonic
off-diagonal block is an $SO(4)$ vector.
Resolving this puzzle was part of the
motivation for this paper.

Another related, unsettled issue is how to obtain
the 32 additional fermions in the heterotic matrix
theory, which is the M(atrix) theory compactified
on $S^1/Z_2$. First it was suggested to add
these fermions by hand to cancel anomalies in
the 1+1 dimensional field theory \cite{KR,BSS1,BM}.
Later Horava \cite{Hor} proposed that they are
zero modes of the off-diagonal blocks that
correspond to 0-8 strings. However, there is
a puzzle of why these fermions are invariant
under surviving supersymmetries. A better
understanding of the 0-8 strings in the M(atrix)
theory should help resolve this problem.
%Another explanation is that they come from
%vanishing two-cycles on $K3$ surface \cite{BR}.

In this paper we study the spectrum of the
off-diagonal blocks in M(atrix) theory that are
supposed to correspond to $p$-$p'$ strings in the
background of a D$p$-brane and a D$p'$-brane.
In particular we show that the spectrum
of zero modes for the off-diagonal blocks matches
the massless spectrum of $p$-$p'$ strings.
Since the string theory results about $p$-$p'$ string
spectrum are most directly seen in the
Neveu-Schwarz-Ramond formalism, while the
M(atrix) description of type \IIA theory
\cite{Motl,BS,DVV} is in the Green-Schwarz
formalism, it is nontrivial to check if their
predictions agree. Moreover, note that D-brane
charges and supersymmetry do not give a complete
characterization for parallel D-brane configurations
in M(atrix) theory. The study of the zero-modes
of  off-diagonal blocks will provide more
information on proper identification of D-brane
backgrounds, and on their dynamical behavior
as well, such as R-R charge and stability etc.

In this paper we will refer to configurations
in M theory by their names in the \IIA theory
that is related to the M theory through
compactification of the (invisible) eleventh
dimension. Hence a D0-brane is a Kaluza-Klein
mode of a graviton, a D2-brane an M-membrane,
a D4-brane a longitudinal M5-brane \cite{Town}.
It is unclear what D6 and
D8-branes in \IIA really correspond to in M theory,
but they are needed to give various D-branes
under compactifications.

We will review related results in string theory in
Sec.\ref{review} and M(atrix) description of
D-brane configurations in Sec.\ref{bundle}.
In Sec.\ref{eom} we derive the equations of motion
for the bosonic and fermionic zero modes
of the off-diagonal blocks, which we will use
to find the zero modes, and explain how to derive their
supersymmetry transformations and low-energy
effective action, for 0-2, 0-4, 0-6 and 0-8 strings
respectively in Sec.\ref{02}-\ref{0608}. Sec. \ref{04}
also includes a discussion on the application of
the off-diagonal zero modes to the matrix description
\cite{BD} of longitudinal fivebranes. More discussions
on the physical implications of our results can be found
in Sec. \ref{0608} and in Sec. \ref{Disc} .

\section{Review of $p$-$p'$ Strings} \label{review}

In this section we briefly review the results in
string theory on $p$-$p'$ strings \cite{Pol1}.
First we consider an open string connecting a
D$p$-brane and a D$p'$-brane parallel to each
other. Since we are using \IIA language,
both $p$ and $p'$ are even integers. Assuming
that $p' \geq p$. In directions $0,1,\cdots,p$,
where the two D-branes overlap, the bosonic fields
$X$ have Neumann boundary conditions on
both ends. In directions $p+1,\cdots,p'$,
they have Dirichlet boundary condition on the
$p$-brane and Neumann condition on the $p'$-brane.
In the rest directions $p'+1,\cdots,9$, the open string
has Dirichlet conditions on both ends.

There will be unbroken supersymmetries for a
system of parallel D$p$-branes and D$p'$-branes,
if and only if the number, $\n$, of the directions
in which the bosonic sector has DN or ND boundary
conditions is 0,4 or 8. (Note that $\n=p'-p$ for a
parallel D$p$- and D$p'$-brane.)

The Ramond sector of the $p$-$p'$ string
has the same kind of boundary conditions
as the bosonic part. It always offers a massless
fermionic $SO(1,9-(p'-p))$ Weyl spinor
(after GSO projection) for the directions with
NN or DD boundary conditions. The NS sector has
the opposite kind of boundary conditions.
Only when $(p'-p)=4$ will there be a massless
$SO(p'-p)$ bosonic Weyl spinor for the directions
with ND or DN boundary conditions.

Since we can always use T-duality to switch a
D$p$-brane to a D0-brane, we only need to
consider four types of open strings: the 0-2, 0-4,
0-6 and 0-8 strings.  In summary, the massless
spectrum for a 0-$p$ string consists of
only a fermionic $SO(1,9-p)$ Weyl spinor,
except that when $p=4$ there is in addition
a bosonic $SO(4)$ Weyl spinor. Below we are
going to verify this spectrum of massless fermionic
and bosonic modes in M(atrix) theory. (Though
it is amusing to note that in M(atrix) theory, the
bosonic off-diagonal blocks that are supposed
to correspond to 0-4 strings are $SO(4)$ vectors!)

\section{M(atrix) Description of D-Brane Configurations}
\label{bundle}

The action of the M(atrix) model is \cite{BFSS}
\be \label{action}
S=\int dt Tr(\frac{1}{4}F_{\m\n}F^{\m\n}
-\frac{1}{2}\Psib\G^{\m}[X_{\m},\Psi]),
\eq
where $\m, \n=0,1,\cdots,9$,
$F_{\m\n}=[X_{\m}, X_{\n}]$
and $X_0=-iD_0=-i(\frac{\del}{\del t}+A_0)$.
$X_{\m}$ and $\Psi_{\a}$ are Hermitian $N\times N$ matrices.
The dynamical and kinematical SUSY transformations
are respectively \cite{BSS}
\bee
\d X_{\m}&=&i\epsb\G_{\m}\Psi, \quad \m=0,1,\cdots,9, \label{dyn1}\\
\d\Psi&=&(D_0 X_i)\G^{0i}\eps+\frac{i}{2}[X_i, X_j]\G^{ij}\eps,
\quad i,j=1,2,\cdots,9, \label{dyn2}
\eqq
and
\be \label{kin}
\d X_{\m}=0, \quad \d\Psi=\tilde{\eps},
\eq
each with 16 generators.

The configuration of a D$p$-brane in M(atrix) theory
is given by big (infinite dimensional) matrices
giving the appropriate $p$-brane charge \cite{BSS}:
\footnote{Only the 2-brane and 4-brane charges are defined
in the SUSY algebra. The 6 and 8-brane charges are
extrapolations of those.}
\be
tr(\eps_{\m_1,\cdots,\m_p}[X_{\m_1},X_{\m_2}]\cdots
[X_{\m_{p-1}},X_{\m_p}]).
\eq
We can choose the $X$'s to satisfy
\be \label{ZZ}
[X_{2n-1},X_{2n}]=F_{(2n-1)(2n)}
\eq
for $n=1,2,\cdots,p/2$
with $F_{\m\n}$ being constant $K\times K$ matrices.
The fermionic partner is taken to be zero.

There are two ways to realize this physical setting
in the M(atrix) theory.
Take the D2-brane as an example.
One way is to set $X_1=R_1 P$, $X_2=R_2 Q$,
where $[P,Q]=i2\pi/N$ \cite{BFSS}.
$P$ and $Q$ can in turn be realized as
$P=-i(2\pi/N)\frac{\del}{\del\s}$ and $Q=-\s$
through an angle parameter $\s\in[0,2\pi)$.
Another way is to first compactify the M(atrix) model
on a torus with radii $R_i, i=1,2$,
and then take the limit $R_i\rightarrow\infty$
if one wishes.
A D$p$-brane configuration corresponds to
a gauge field configuration with certain
topological charge \cite{BSS}
(the $k$-th Chern character $Q_k=\int trF^k$ for $k=p/2$.)
on the dual torus which becomes infinitesimal
in the large radii limit.
The $X$ matrices in the $p$-brane directions
become ($(-i)$ times) the covariant derivatives.
For a D2-brane it can be taken as, say,
$X_1=-i2\pi R_1\frac{\del}{\del\s_1}$
and $X_2=-i2\pi R_2\frac{\del}{\del\s_2}-R_2\s_1$.

For our purpose the difference between the two descriptions
is only a scaling in the derivatives.
For simplicity in notation
we choose to use the latter description in this paper.

A static D$p$-brane configuration (\ref{ZZ})
preserves half of the total SUSY if and only if
the $F$'s are proportional to the unit matrix,
in which case 16 linear combinations of
the dynamical and kinematical SUSY are preserved \cite{BSS}.
These states contain D0, D2,..., D$(p-2)$-branes
in addition to the D$p$-brane.
The kinematical SUSY (\ref{kin}) is never preserved by itself.
The condition for part of the dynamical SUSY to be preserved is
\be \label{FF}
F_{12}-\sum_{i=2}^{p/2}\ve_i F_{(2i-1)(2i)}=0
\eq
for some $\ve_i=\pm 1$.
It preserves $1/2^{(p/2-1)}$ of the dynamical SUSY
parametrized by $\eps$ satisfying
\be
\G^{12}\G^{(2i-1)(2i)}\eps=\ve_i\eps, \quad i=2,3,\cdots,p/2.
\eq
Because $tr(F_{12}^2)\neq 0$ it follows from (\ref{FF})
that any D6 or D8-brane configuration with unbroken dynamical SUSY
must always include D4-branes.
A discussion on general bound states from
the low energy D-brane point of view can be found in \cite{GR}.

If all $F_{\m\n}$'s in (\ref{ZZ})
are proportional to the unit matrix,
they define a natural complex
structure on the dual torus.
It can be used to view the dual torus $T^p$ as
composed of $p/2$ complex tori $T^2$.
A D$p$-brane with unit $p$-brane charge can be realized by
a $U(K)$ gauge field with twisted boundary conditions.
This is analogous to how one defines
a long string \cite{DVV,BS}
in the conjugacy class of length $K$.
The unit D$p$-brane charge means a twisted bundle
with the minimal topological charge on each $T^2$.

An explicit construction of
the minimal twisted bundle
of the fundamental representation of $U(K)$
is given in Ref.\cite{GRT}.
There the gauge fields can be chosen as $A_1=0$
and $A_2=-i(\s_1/2\pi K)\one$,
where we use $\s_i$ as the coordinates on $T^2$
normalized to range between $0$ and $2\pi$
and $\one$ is the unit matrix.
The field strength $F_{12}$ is then $1/2\pi K$.
The quasi-periodic boundary conditions on $A$ are \cite{GRT}:
\bee
&A_{\m}(2\pi,\s_2)=\O_1(\s_2)A_{\m}(0,\s_2)\O_1^{-1}(\s_2)
+\O_1(\s_2)\del_{\m}\O_1^{-1}(\s_2), \\
&A_{\m}(\s_1,2\pi)=\O_2(\s_1)A_{\m}(\s_1,0)\O_2^{-1}(\s_1)
+\O_2(\s_1)\del_{\m}\O_2^{-1}(\s_1),
\eqq
where $\O_1$ and $\O_2$ can be chosen as
\be
\O_1(\s_2)=q^{\s_2/2\pi} U, \quad \O_2(\s_1)=V,
\eq
where $q=e^{i2\pi/K}$, $U_{ij}=q^{i}\d_{ij}$ and
$V_{ij}=\d_{i+1,j}$ with $i,j=0,\cdots,K-1\; (mod \; K)$.
$U$ and $V$ satisfy $UV=q^{-1}VU$.
It can be checked that $\O_1(2\pi)\O_2(0)=\O_2(2\pi)\O_1(0)$.
This is in contrast with the twisted bundle of $SU(K)/\Z_K$
where one has $\O_1(2\pi)\O_2(0)=\O_2(2\pi)\O_1(0)Z$
for some element $Z$ in the center $\Z_K$ of $SU(K)$ \cite{tH}.

The bundle in fundamental representation has the corresponding
boundary conditions:
\be
\phi(2\pi,\s_2)=\O_1(\s_2)\phi(0,\s_2), \quad
\phi(\s_1,2\pi)=\O_2(\s_1)\phi(\s_1,0).
\eq
Note that consistency of the boundary conditions
requires
$$\O_1(2\pi)\O_2(0)=\O_2(2\pi)\O_1(0).$$
A section of the bundle has the general form of \cite{GRT}
\be \label{section}
\phi_j(\s_1,\s_2)=\sum_{m\in\Z}
\phih(\s_2/2\pi+j+mK)q^{(\s_2/2\pi+j+mK)\s_1/2\pi},
\quad j=0,1,\cdots,K-1,
\eq
for an arbitrary function $\phih$ for which the series converges.

Since this is the D-brane analogue of a long string in
the conjugacy class of length $K$,
this gauge field configuration
is identified with a single D2-brane
instead of $K$ D2-branes.
Here $K$ gets interpreted as the longitudinal
momentum carried by the single D2-brane, as can be
seen by examining its light-cone energy.

It is essential that the gauge group is $U(K)$
instead of $SU(K)$.
Although there are twisted $SU(K)/\Z_K$ bundles
in the adjoint representation \cite{tH}
with the same topological charge,
there is no corresponding vector bundle in
the fundamental representation,
because the element $Z$ acts nontrivially on
the fundamental representation while
it acts trivially on the adjoint.
Note that the presence of anything other than the D4-branes
introduces off-diagonal blocks in
the fundamental representation.
Hence, for instance,
although one can use two copies of
the twisted $SU(2)$ bundle on $T^2$
with (anti-)self-duality
to construct pure D4-brane states
preserving half of the dynamical SUSY,
at this moment it is unclear how to describe their
interaction with other D-branes.

\section{Equations of Motion} \label{eom}

Consider a D$0$-brane very close to a D$p$-brane.
We decompose the matrix fields into the block form:
\be \label{block}
X_{\m}=\left(\begin{array}{cc}
                Z_{\m} & y_{\m} \\
                \yd_{\m} & x_{\m}
             \end{array}\right); \quad
\Psi=\left(\begin{array}{cc}
              \Th & \th \\
              \thd & \psi
           \end{array}\right),
\eq
where $Z_{\m}$ represents the D$p$-brane
and $x_{\m}$ the D$0$-brane.
The generalization to many D$p$-branes and D0-branes
is straightforward.
While the $Z$'s are realized as covariant derivatives,
the $x$'s in general can have nontrivial
coordinate dependence on the dual torus,
but when we take the limit of infinite radii,
only coordinate-independent states can have
finite energy and remain coupled to the theory.
(We allow infinite energy for $Z$ because
it is just the energy for the D$p$-brane.)

For simplicity we choose the coordinates of spacetime
such that $x_{\m}=0$ and $Z_a=0, a=p+1,\cdots,9$.
The D$p$-brane is parallel to the directions $1,2,\cdots,p$
and the D$0$-brane is right on top of it.
The diagonal part is taken as the background configuration.

When putting the two D-branes together
as in (\ref{block})
and set the off-diagonal parts to be zeros,
one can check easily that part of the supersymmetry is
preserved only if $p$ is $0,4$ or $8$.

To count the number of zero modes,
or equivalently to count the dimension of
the moduli space for this background,
it is easier to consider the perturbation of this
background and keep only the lowest order terms
to obtain linear differential equations for the
perturbative fields $y$ and $\th$.
In this way we count the dimension of the tangent space
on the moduli space.
One may also introduce perturbations in
the diagonal blocks for fluctuations on the D$p$-brane
and deviations of the D0-brane from the origin,
but here we are for the time being only
interested in the off-diagonal blocks $y$ and $\th$
since they represent the $p$-$p'$ strings.
The perturbations of the diagonal blocks can be studied
in the same way we study the off-diagonal part.
To the lowest order in perturbation,
the perturbative diagonal and off-diagonal parts
are not correlated,
hence we can treat the off-diagonal ones alone.

Plugging the expression of the matrix fields
(\ref{block}) into the action of the M(atrix) model (\ref{action}),
we find
\be
S=\int dt (L_Z+L_x+L_y),
\eq
where $L_Z$ and $L_x$ are of the same form as (\ref{action})
except that we replace ($X,\Psi$) by ($Z,\Th$) and ($x,\psi$),
respectively.
$L_y=L_B+L_F$ with
\bee
L_B&=&\frac{1}{2}|(Z_{\m}y_{\n}-Z_{\n}y_{\m}+y_{\m}x_{\n}-y_{\n}x_{\m})|^2
+y_{\m}^{\dagger}[Z_{\m}, Z_{\n}]y_{\n}
-[x_{\m}, x_{\n}]y_{\m}^{\dagger}y_{\n}\nn\\
&&-y_{\m}^{\dagger}y_{\n}y_{\m}^{\dagger}y_{\n}
+\frac{1}{2}y_{\m}^{\dagger}y_{\m}y_{\n}^{\dagger}y_{\n}
+\frac{1}{2}y_{\m}^{\dagger}y_{\n}y_{\n}^{\dagger}y_{\m}
\eqq
and
\be
L_F=-\th^{\dagger}\G^0\G^{\m}(Z_{\m}\th-\th x_{\m})
+\th^{\dagger}\G^0\G^{\m}(\Th y_{\m}-y_{\m}\psi)
+(y_{\m}^{\dagger}\Th-\psi y_{\m}^{\dagger})\G^0\G^{\m}\th.
\eq
For more than one D0-branes the $x$'s are matrices
and we need to take traces for these formulas.

>From the action
one can derive the equations of motion for $y$ and $\th$.
Since the Hamiltonian for a time-independent background
in the temporal gauge ($A_0=0$) is minimized
by time-independent $y$,
we look for time-independent solutions for $y$ and $\th$.
Ignoring the time derivative and higher order terms,
we find
\bee
&D^{\m}(D_{\m}y_{\n}-D_{\n}y_{\m})+[D_{\n}, D_{\m}]y^{\m}=0,
\quad &\m, \n=1,\cdots,p,\label{eom1-y}\\
&D^{\m}D_{\m}y_a=0, \quad &a=p+1,\cdots,9,\label{ya}
\eqq
where $D_{\m}=iZ_{\m}$ are
covariant derivatives on the dual torus $T^p$ as
$D_{\m}=2\pi R_{\m}(\frac{\del}{\del\s^{\m}}+A(\s))$,
$\m=1,\cdots,p$.
Eq.(\ref{eom1-y}) has to be supplemented by
the gauge-fixing condition
\be \label{eom2-y}
D_{\m}y^{\m}=0.
\eq
Using (\ref{eom2-y}), eq.(\ref{eom1-y}) can be written as
\be \label{eom3-y}
D_{\m}D^{\m}y_{\n}+2[D_{\n},D_{\m}]y^{\m}=0.
\eq

The equation of motion for $\th$ is
\be\label{eom-th}
\G^{\m}D_{\m}\th=0.
\eq

In terms of the covariant exterior derivative $d_A$,
its dual $d_A^*$, the Hodge dual $*$
and the projection $P=\frac{1}{2}(1-*)$ (so $P^2=P$),
eqs.(\ref{eom1-y}) and (\ref{eom2-y}) now read
\bee
& d_A^* Pd_A y=0, \\
& d_A^* y=0,
\eqq
where $y=y_{\m}d\s^{\m}$.
These equations are formally the same as those
for the instanton zero modes,
which correspond to perturbations
of the $Z$'s above.
The only difference is that the perturbations of $Z$
is in the adjoint representation of $U(K)$,
while $y$ is in the fundamental representation.

Because we are considering the Euclidean torus,
the inner product $\la\cdot|\cdot\ra$
defined by integration on the torus
and the trace of matrices is positive definite.
Hence $\la y|d_A^* Pd_A y\ra=0$ implies that
\be \label{PdAy}
Pd_A y=0.
\eq
In addition, eq.(\ref{ya}) implies that
$\la D_{\m}y_a|D^{\m}y_a\ra=0$
and so $D_{\m}y_a=0$,
which means that the topological charge vanishes
unless $y_a=0$.
Thus we conclude that $y_a=0$ for $a=p+1,\cdots,9$.

\section{0-2 Strings} \label{02}

Let $Z_1$ and $Z_2$ be realized as $U(K)$ covariant derivatives
on the dual torus as $Z_i=-i D_i$ with
$D_i=\frac{\del}{\del\s_i}+A_i$ as given in Sec.\ref{bundle}
so that
\be
[Z_1,Z_2]=if\one,
\eq
where $f=2\pi R_1 R_2/K$.
For simplicity we are considering unslanted torus
with radii $R_1=R_2=1/2\pi$.
It is straightforward to generalize to slanted tori
with arbitrary radii.
Let $z=(\s_1+i\s_2)/2\pi$, $\zb=(\s_1-i\s_2)/2\pi$
be the complex coordinates on $T^2$,
and let $\cD=D_1-iD_2$ and $\cDb=D_1+iD_2=-\cD^{\dagger}$,
then $[\cD,\cDb]=2f\one$.
It follows that
\be
D^2=\cD\cDb-f=\cDb\cD+f,
\eq
where $D^2=D_1^2+D_2^2$.
Note that the algebra of $\cDb$ and $-\cD$
is the canonical commutation relation
for annihilation and creation operators scaled by $2f$.
Therefore the spectrum of $\cD\cDb$ is $\{0,-2f,-4f,\cdots\}$
and the spectrum of $D^2$ is
\be \label{spec}
\{-f,-3f,-5f,\cdots\}.
\eq

The fermionic zero modes satisfy (\ref{eom-th}),
which gives $(D_1+\G^1\G^2 D_2)\th=0$, so that
\be
\cD\th_+=0, \quad \cDb\th_-=0,
\eq
where $\th_{\pm}$ are the two Weyl components of $\th$
satisfying $i\G^1\G^2\th_{\pm}=\pm\th_{\pm}$.
Because $\la\th_+|\cDb\cD\th_+\ra=\la\th|(D^2-f)\th\ra<0$
for any $\th_+\neq 0$,
we must have $\th_+=0$.
The solution of $\th_-$ is obviously the vacuum state
annihilated by $\cDb$.

One can easily get the explicit expression of the vacuum
as a section of the twisted bundle
using the explicit construction in Sec.\ref{bundle}.
Another way is to note that
the equation $\cDb\phi=0$ has the general solution of
$\phi =\exp (-\frac{\pi}{4K}(\bar{z}^2+2z\bar{z}))f(z)$,
where $f(z)$ is an arbitrary holomorphic function.
For $\phi$ to be a section of the twisted bundle,
we need to impose the quasi-periodic
boundary conditions on $\phi$.
One then sees that
$f(z)$ is related to the third elliptic theta function $\vartheta_3$ and
the solution is
\be \label{theta}
\phi_k(\s_1,\s_2)=\exp\{\pi/K[2i(\s_1/2\pi)(\s_2/2\pi+k)-(\s_2/2\pi+k)^2]\}
\vartheta_3(q|\pi(z+ik)),
\eq
where $q=\exp (-\pi K)$ and $\phi_k$ ($k=0,1,\cdots,K-1$)
gives a section on the vector bundle in the fundamental representation.
Applying the creation operator $\cD$ to the vacuum
one obtains other eigenstates of the operator $D^2$.

Obviously the zero mode of $\th_-$ is just given by
the solution (\ref{theta}).
The fermionic zero mode is an $SO(2)$ Weyl spinor
with negative chirality.

The equations of motion (\ref{eom3-y}) for $y_{\m}$ are
\be
(D^2-2f)y=0, \quad (D^2+2f)\yb=0,
\eq
where $y=y_1+iy_2$, $\yb=y_1-iy_2$.
The constraint (\ref{eom2-y}) is
\be
\cD y+\cDb\yb=0.
\eq
Since the spectrum of $D^2$ is given by (\ref{spec}),
we see that there is no solution for $y$, $\yb$,
hence there is no bosonic zero mode.

\section{0-4 Strings} \label{04}

We decompose the 10 dimensional $\g$-matrices as
\bee
\G^0&=&i\s_2\otimes\one\otimes\one, \\
\G^{\m}&=&\s_1\otimes\g^{\m}\otimes\one, \quad \m=1,\cdots,4,\\
\G^a&=&\s_1\otimes\g^5\otimes\g^a, \quad a=5,\cdots,9, \\
\G^{10}&=&\s_3\otimes\one\otimes\one,
\eqq
where the $\s_i$'s are the Pauli matrices satisfying
$\s_1\s_2=i\s_3$, the $\g^{\m}$'s ($\m=1,\cdots,4$)
are the $\g$-matrices
for $SO(4)$, the $\g^a$'s ($a=5,\cdots,9$) are
$SO(5)$ $\g$-matrices.
Corresponding to this decomposition,
a 10-dimensional spinor $\th_{i\a\b}$ has three indices,
where $i=\pm$, $\a,\b=1,\cdots,4$.
For Weyl spinors with positive (negative) chirality
one has $i=+$ ($i=-$).
Since all spinors in this theory are
10-dimensional Weyl spinors with positive chirality,
we will omit the index $i$ in the following.
We consider the case where the gauge field for
the D4-brane background is self-dual,
so that half of the dynamical SUSY
with parameter $\eps$ satisfying
\be \label{eps4}
\G^1\G^2\G^3\G^4\eps=\eps
\eq
is preserved.

The number of zero modes for the case of $T^4$
for (anti-)self-dual gauge field configurations
can be determined using
the index theorem \cite{ASch,Baal}.
The number of spinorial zero modes is found to be $\a_V k$,
where $\a_V$ is the Dynkin index for the representation $V$
of the fermions,
and $k$ is the instanton number.
The number of vectorial zero modes is $2\a_V k$.
For any $U(K)$, the number of
spinorial and vectorial zero modes
in the fundamental representation
are $k$ and $2k$, respectively.
The general formula \cite{ASch}
for spinorial and vectorial zero modes in
arbitrary representation $V$ on a 4-manifold $M$ are
$\a_V k+\frac{1}{8}dimV\tau(M)$ and
$2\a_V k-\frac{1}{2}dimV(\chi(M)+\tau(M))$,
respectively, where
$\tau(M)=\frac{1}{96\pi^2}\int\eps^{\m\n\a\b}R_{\m\n\l\r}R_{\a\b}^{\l\r}dv$
is the signature of $M$ and
$\chi(M)=\frac{1}{128\pi^2}\int\eps^{\m\n\a\b}\eps^{\l\r\g\d}
R_{\m\n\l\r}R_{\a\b\g\d}dv$ is the Euler characteristic.

While according to the the index theorem
the number of zero modes is independent of
the details of
(anti-)self-dual gauge field configurations,
here we give for example an explicit construction of a twisted bundle
for the cases with $R_1 R_2=R_3 R_4$.
On each $T^2$ factor of $T^4$, one can construct
a twisted $U(K)$ bundle as in Sec.\ref{bundle}.
When putting them together, we obtain a $U(K^2)$ bundle
with unit instanton number:
$\frac{1}{8\pi^2}\int tr(F^2)=1$.
Unlike the case of twisted $SU(K)/\Z_K$ bundles which
can have fractional instanton numbers,
for $U(K)$ the instanton number are always integral \cite{GR}.
A section on the twisted $U(K^2)$ bundle on $T^4$ has
the general form of a linear combination of
products of sections on each $T^2$:
$\phi_j(\s_1,\s_2)\phi_k(\s_3,\s_4)$,
where $\phi$ is defined by (\ref{section})
for $j, k=0,1,\cdots,(K-1)$.
Indices $j$ and $k$ compose an index for
the fundamental representation of $U(K^2)$.
In general one can also consider a $U(K_1)$ and $U(K_2)$ bundle
on the two $T^2$ factors, respectively,
and obtain a $U(K_1 K_2)$ bundle on $T^4$.

Because the supersymmetry is not completely broken,
the solution of fermionic zero modes can be used to
obtain the solution of bosonic zero modes.
The solution of the fermionic and bosonic zero modes
can be obtained explicitly by considering
$T^4$ as $T^2\times T^2$ and using the methods in Sec.\ref{02}.
Let the $SO(4)$ spinor satisfying (\ref{eom-th})
be denoted by $\th^0$.
It is easy to see that the fermionic zero mode satisfies
$i\G^1\G^2\th^0=i\G^3\G^4\th^0=-\th^0$,
which implies that $\th^0$ is of negative chirality
as an $SO(4)$ Weyl spinor:
$\G^1\G^2\G^3\G^4\th^0=-\th^0$.
(If the gauge field is anti-self-dual,
the zero mode will be a Weyl spinor with positive chirality.)
For a single D4-brane there is only one fermionic zero mode,
which is given by the product of the solutions (\ref{theta})
on each $T^2$ factor in $T^4$.

The general solution of fermionic zero modes
can then be written as $\th_{\rd\b}=\th^0_{\rd}\chi_{\b}$
(and $\th_{\r\b}=0$),
where $\rd=1,2$ ($\r$) is the index for an $SO(4)$ Weyl spinor
with negative (positive) chirality
and $\chi$ is an $SO(5)$ spinor for the directions
$5,\cdots,9$.
Thus $\chi$ represents the massless fermionic mode
from the Ramond sector of the 0-4 string.

Since the equations of motion for $y$ and $\th$
are supersymmetric,
the bosonic zero mode can be obtained by
SUSY transformation \cite{Zum} as
\be
y^{\m}=iv^{\dagger}_{\r}\g^{\m}_{\r\rd}\th^0_{\rd},
\eq
where $v_{\r}$ is an $SO(4)$ Weyl spinor
with positive chirality.
This comes from the SUSY transformation of $y$:
$\d y_{\m}=i\bar{\eps}\G_{\m}\th$.
When one replaces in this transformation $\th_{\a\b}$ by
the zero mode $\th^0_{\rd}$,
$\d y$ will satisfy the equations of motion of $y$
for any $\eps_{\r}$ in the SUSY preserved by
the background (\ref{eps4}).
It follows that the $y$ given by the above expression
is a zero mode of $y$.
Since $\th^0$ is a function (bosonic),
$v$ is a bosonic variable.
It matches the massless bosonic field from the NS sector
of the 0-4 string.
Here it is amusing to see how supersymmetry dictates
the zero modes of a field $y$ in vector representation
to be described by a variable $v$ in spinor representation.
The index theorem \cite{ASch} assures
us that these
are all the zero modes in the theory,
giving precisely the massless spectrum of 0-4 strings.
The supersymmetry transformation between $\chi$
and $v$ is induced from the SUSY transformation
between $\th$ and $y$ by factoring out the common factor
of $\th^0$.
Up to first order perturbation,
the SUSY transformation of $\th$ is:
$\d\th=\frac{1}{2}(D_{A}y_{B}-D_{B}y_{A})\G^{AB}\eps$,
where $A,B=0,1,\cdots,9$.
Using (\ref{PdAy}), (\ref{ya}) and (\ref{eps4}),
one finds
\footnote{Our notation is slightly different from
that of Ref.\cite{BD}.}
\bee
& \d v_{\r}=\chi^{\dagger}_{\a}\eps_{\r\a}, \\
& \d \chi_{\a}=2i(D_0 v^{\dagger}_{\r})\eps_{\r\a}.
\eqq

The instanton connection lies in $SU(2)_R\in SO(4)$ which
is supposed to be the global R-symmetry for the action of 0-4 strings.
Field $v$ carries the fundamental index of $SU(2)_R$.
Let $\tau^i$ denote generators of the R-symmetry group. There are two
possible $SU(2)_R$ invariant D-terms, $\sum_i |v^+\tau^i v|^2$ and
$|v^+v|^2$. The two terms are different when there are more than one
D0-branes, in which case only the first is actually present
in the action \cite{Pol1}.
These D-terms are expected to arise from the $F^2$ term in
the Super Yang-Mills theory. Upon expanding this term in $y$ one
finds $tr |y_\mu y_\nu^+ -y_\nu y_\mu^+|^2$ and $|y_\mu^+y_\nu
-y_\nu^+y_\mu|^2$. For a given instanton background, since $SU(2)_R$
is broken explicitly, these terms do not give those $SU(2)_R$
invariant D-terms. Only after averaging over the moduli space does
one expect that the symmetry $SU(2)_R$ is restored. However, we do not
know how to rule out the $U(1)$ D-term $|v^+v|^2$.

The above discussion easily generalizes to the case of instanton
number $k$. There are $2k$ zero modes for $y_\mu$, and can be
interpreted as the fundamental of $U(k)\times SU(2)_R$, where
$U(k)$ is the gauge group associated to $k$ coincident D4-branes.

In ref. \cite{BD}, an action describing M(atrix) theory
of a longitudinal 5-brane is proposed. Since a longitudinal
5-brane in M-theory corresponds to a D4-brane in type \IIA
string theory, some extra dynamical variables corresponding
to 0-4 strings were needed and were introduced by hand.
Their quantum numbers are exactly the same as those of the
variables $v$ and $\chi$ that we have discussed above.
Thus, it is natural to identify the additional variables
introduced by Berkooz and Douglas \cite{BD} with the degrees
of freedom associated with the off-diagonal zero modes. We have
verified that indeed the action of the latter naturally derives
from the fundamental M(atrix) model action, and it agrees with
the action postulated in ref. \cite{BD}, with a possible
$U(1)$ D-term as we mentioned above. (In the derivation,
the coefficient of each term in the action is determined by
an integral of a product of the zero mode solutions $\th^0$.
We have not been able to calculate all coefficients; presumbly
they are uniquely determined by the surviving supersymmetry.)

In addition to the variables $v$ and $\chi$, the action
in ref. \cite{BD} has included also fields describing
fluctuations of the longitudinal fivebrane background,
which in our approach correspond to fluctuations residing
in the diagonal blocks. In principle one can consider
fluctuations of all blocks in the matrix fields for a
given background, and then solve the exact (nonlinear)
equations of motion. The parameters analogous to $v$ and
$\chi$ above for the general solutions
correspond to the massless modes of
the whole system of ($p'$-$p$)-branes.
In the above we have only solved the linearized
equations of motion for the off-diagonal blocks.
The supersymmetry derived from our solutions will only
hold to the lowest order in perturbation.
If one solves the exact nonlinear equations of motion,
one should be able to derive the exact SUSY transformation
among the zero mode parameters.

In the above we have only considered the case with
vanishing distance between the D0-brane and the D4-brane.
When we pull the D0-brane away from the D4-brane,
the zero modes will gain masses proportional to the distance.
But we expect that the number and representation
of the lowest energy modes will remain the same
as the zero modes. The proposal of Ref.\cite{BD}
contains only the lowest energy modes and
therefore should be viewed as a low energy effective theory.

\section{0-6 Strings and 0-8 Strings} \label{0608}

The case of 0-6 strings and 0-8 strings
can be studied in a similar fashion
as the 0-2 and 0-4 strings.
To generalize the consideration for 0-2 and 0-4 strings
to 0-$p$ strings for $p=2,4,6,8$,
we choose the gauge field configuration for the D$p$-brane
to be $p/2$ copies of the gauge field configuration
on $T^2$ described in Sec.\ref{02},
that is,
\be
[Z_{2i-1},Z_{2i}]=if\one, \quad i=1,\cdots,p/2,
\eq
where $f=1/2\pi K$.
This defines a twisted $U(K^{p/2})$ bundle with
unit $p$-brane charge:
$\frac{1}{k!(2\pi)^k}\int tr(F^k)=1$ for $k=p/2$.

We focus our attention on the first copy of $T^2$.
Let $y=y_1+iy_2$ and $\yb=y_1-iy_2$.
The equations of motion for them are
$(D^2-2f)y=0$ and $(D^2+2f)\yb=0$,
where $D^2=\sum_{\m=1}^{p}D_{\m}^2$ for a D$p$-brane.
Since the spectrum of $D_1^2+D_2^2$ is shown to be
$\{-f,-3f,-5f,\cdots\}$ in Sec.\ref{02},
the spectrum of $(D^2+2f)y$ is
$\{-(p/2-2)f,-p/2f,\cdots\}$ and
the spectrum of $(D^2-2f)$ is purely negative for any $p$.
It then follows that there is a zero mode for $y$
only if $p=4$.

The equation of motion for the fermionic mode
is decomposed into $p/2$ equations for a D$p$-brane:
$(D_{2i-1}+\G^{2i-1}\G^{2i}D_{2i})\th=0$, $i=1,\cdots,p/2$.
Obviously the solution of $\th$ is simply
the product of the solution (\ref{theta}) for each
copy of $T^2$ and
it is of negative chirality on each $T^2$
so that $\G^1\cdots\G^p\th=i^{p/2}\th$.
The index theorem \cite{AS}
\be
ind(E,D)=(-1)^{m(m+1)/2}\int_M ch(\oplus_r(-1)^r E_r)
\left.\frac{Td(TM^{\C})}{e(TM)}\right|_{vol}
\eq
can be used to show that there is
only one fermionic zero mode if one can show that
there is no zero mode of the opposite chirality:
$\G^1\cdots\G^p\th=-i^{p/2}\th$.
Indeed one can consider the spectrum of the Dirac operator
squared $(\G^{\m}D_{\m})^2=D^2+\G^{\m\n}[D_{\m},D_{\n}]$.
The spectrum of $D^2$ is given above and the spectrum
of the second term is $\{-(p/2)f,-(p/2-2)f,\cdots,(p/2)f\}$.
It follows that the zero mode must have negative
chirality on each $T^2$.

The result is therefore that for a 0-$p$ string
there is always a single fermionic zero mode
and there is no bosonic zero mode except for the 0-4 string.
This is in agreement with the results of string theory.

In Sec.\ref{04} we showed that the SUSY property of
the zero modes of a 0-4 string
follows from that of the off-diagonal blocks.
The SUSY transformation of the zero mode for a 0-8 string
can also be derived from the SUSY of SYM.
Now let us show that the bosonic zero modes derived
from the fermionic zero modes using the SUSY transformation
as in Sec.\ref{04} merely vanish.
Note that the $SO(1,9)$ symmetry is decomposed into
$SO(1,1)\times SO(4)\times SO(4)$ for the 0-8 string,
where the D8-brane has two D4-branes with it.
The $\G$ matrices can be taken as in Sec.\ref{04}.
A 10-dimensional spinor $\th_{\pm\a\b}$
has three indices corresponding to the three factors
of orthogonal group.
The SUSY preserved by the D8-brane background
is parametrized by $\eps$ with positive or negative chirality
on both factors of $SO(4)$;
and the zero mode of $\th$ has negative chirality
for both $SO(4)$.
Since a given $\G$-matrix can change the chirality of only one
of the two copies of $SO(4)$,
the SUSY transformation $\d y_{\m}=i\bar{\eps}\G_{\m}\th$
vanishes for $\th$ being the zero mode
and does not give nontrivial solutions to $y$.

It is easy to see that the fermionic zero mode is given by
$\th_{+\sd\rd}=\chi_+
\lam^0_{\sd\rd},$
where the $\lam^0$ is the zero mode solution on $T^8$.
%which is simply the product of
%two factors of $\th^0$ correspond to
%the zero modes on the two factors of $T^4$
%(which are products of the theta function (\ref{theta})
%on each $T^2$ factor of the $T^4$),
%with all other components of $\th$ vanishing.
The SUSY transformation of the fermionic zero mode
is trivial ($\d\chi=0$) because all $y$'s vanish.
This agrees with the proposal of Horava \cite{Hor}
to interpret the zero modes as the extra fermions
needed in the heterotic matrix theory \cite{DF,KR,BM}.

If the gauge field configuration for a D4-brane is not
(anti-)self-dual,
it is found that \cite{Baal} the configuration is not stable
because of the existence of negative energy states
in the perturbation of the gauge fields.
Therefore all states tend to decay into an (anti-)self-dual state
with the same topological charge.
In our consideration of the off-diagonal blocks $y$,
the spectrum of the operator $(-D^2\pm 2f)/2$
corresponds to the energy of states on the 0-$p$ strings.
For the 0-2 string the lowest energy of $y$ is $-f<0$
and it signifies the instability of the system.
This is consistent with the fact that the 0-brane
tends to distribute uniformly
over the D2-brane \cite{Tay1} to form a bound state.
For D4-branes corresponding to (anti-)self-dual
configurations the lowest energy of $y$ is 0,
but otherwise there would be states with negative energy
equal to $-|f_1-f_2|$ where $F_{12}=if_1$ and $F_{34}=if_2$.
In general for a 0-$p$ string the lowest energy is
the minimum of $\{(\sum_{i=1}^{p/2}f_i-2f_j)/2\;|\;j=1,\cdots,p/2\}$.
While there are D2-branes inside the D4, D6 and D8-brane
configurations we considered,
the interaction between the D$p$-brane and D0-brane
includes the attraction from the D2-brane and
repulsion from the D6.
(The D0-brane is marginally bound to a pure D4-brane.) \cite{ASJ}
If the lowest energy is positive, zero or negative,
it means that the configuration is stable, marginally stable
or unstable, respectively. In the cases of D6 and D8-branes,
the negative modes are due to the D2-branes inside the higher
branes. Take D6-brane as an example. Let $f_i>0$ and
$f_1=f_2$, then there is a D4-brane wrapping around the
first two tori. If $f_3>2f_1$, there is a negative mode
of energy $2f_1-f_3$. Apparently, the attractive force due to
the D2-brane on the third torus overcomes the repulsive force
of the D6-brane.

Generically for D$p$-branes there is a Fock space ${\cal H}_i$
for each $T^2$, where $-\cD_i/\sqrt{2f_i}$ and $\cDb_i/\sqrt{2f_i}$
act as the creation and annihilation operators.
After imposing the constraint (\ref{eom2-y}),
the spectrum of $y_{\m}$ is found to be
\be
\{(\sum_{j=1}^{p/2}f_j-2f_i)/2,\; (\sum_{j=1}^{p/2}(2n_j+1)f_j+2f_i)/2\;|
\; i=1,\cdots,p/2;\; n_j=0,1,2,\cdots\}.
\eq

\section{Discussions} \label{Disc}

In this paper, we have presented a general framework
and a systematic analysis for the zero modes in the
off-diagonal blocks in M(atrix) theory. More concretely,
we have shown how to determine the number of zero
modes by index theorem and surviving supersymmetry,
and moreover we have determined the quantum numbers
of the zero modes, including the chirality of the
fermion zero modes. These quantum numbers are nontrivial,
and crucial for us to show the agreement with string
theory predictions on open $p-p'$ strings stretching
between D-branes, providing one more check for
M(atrix) theory. Previously in Refs.
\cite{Aharony,Lif1,Lif2}, in the middle of computing
the effective potential between a D0- and Dp-brane,
the energy levels of the off-diagonal block have
been determined using a slightly different
representation for the Dp-brane. But the zero
modes were not mentioned and identified, and their
quantum numbers were not studied.

Now let us discuss the significance in M(atrix) theory
of the zero modes residing in the off-diagonal blocks.
First we have shown in Sec. \ref{04} that for the case
of a longitudinal fivebrane, the degrees of freedom
associated with the off-diagonal zero modes naturally
provide the extra degrees of freedom put in by hand by
Berkooz and Douglas, ref. \cite{BD}. And we have
checked that the action they postulated are derivable
from the M(atrix) theory action, with a possible
$D$-term. Indeed, in this case, besides the right
topological number (or brane charge), the correct
counting of zero modes we found in Sec. \ref{04}
is crucial for justifying our identification
of a longitudinal 5-brane with proper instanton
configuration on $T^4$ rather than on $S^4$. Also
the correct number of zero modes is crucial for
a check of the correct tension and R-R
charge for the longitudinal 5-brane. It
is argued in \cite{BD} that upon integrating
out 0-4 strings the long range force
between a longitudinal 5-brane and a
probe supergraviton is generated.
If we had a different number of zero modes
we would obtain a gravitational field with a
different magnitude for the 5-brane. Also
as shown in Ref. \cite{BD}, the R-R charge
of a longitudinal 5-brane manifests itself in the
Dirac quantization of a membrane moving
in its background.
By realizing the membrane as a collection of
D0-branes, the zero modes on the 0-4 strings
would induce fields on the membrane.
It is the fermion zero
mode $\chi$ induced on the membrane
that is responsible for generating the
Berry phase.
In fact, by T-duality the induced zero mode
is related to the zero mode on a 0-6 string.
Our results in Sec.\ref{0608} provide the proof
for the existence of a single chiral zero mode
necessary for the correct Berry phase.
Had we had two zero modes,
we would have generated twice the correct
Berry phase, and therefore twice the R-R charge.
As pointed out in Sec.6, in the background of
$k$ instantons, there are $k$ fermionic zero
modes. The Berry phase is then $k$ times
as large, and this signals that there are $k$
units of R-R charge.

Upon compactifying on a 5-torus $T^5$, instanton
strings will appear in the spectrum. These are
part of constituents of some 5D black holes
\cite{LM,DVVa}. A 5D black hole is described
by a long instanton-string carrying momentum.
Probing the black hole with a supergraviton,
one expects that the corresponding static potential
as well as the velocity dependent force are generated
by integrating out the off-diagonal blocks. This is
shown to leading order in Ref. \cite{JM}, where
the full $5+1$ massive modes are integrated out.
It is an interesting question whether the relevant
terms can be generated by integrating out only the
zero modes discussed in Sec.\ref{04}.

It should be interesting to compare our result with
the work of Ref.\cite{Doug}. There a D5-brane is
interpreted as an instanton inside 9-branes. The
probe is a D1-brane. The 1-5 string sector is
constructed with the D-brane technology.
A (0,4) sigma model in an instanton background
\cite{Witten} results from integrating out the
massive 1-5 strings. There are two differences
between the case under discussion and Ref.
\cite{Doug}. First, it is crucial for us to work
with $T^4$, only then we have the correct
number of zero modes. Second, the $SU(2)_R$
symmetry in our problem comes from the $SO(4)$
of $T^4$, while the $SU(2)_R$ of \cite{Doug}
does not act on the gauge field, since the gauge
field carries an index transverse to the D5-brane.

Finally, the origin of $p-p$ strings is also easy
to see. When $p=2$, the world-volume action
is written down \cite{BSS}. For $p=4$, one can
consider zero modes of the fundamental of
$SU(2)\times SU(2)\in SU(4)$ in the background
instanton number 2 solution with a gauge group
$SU(4)$. It is important to embed the instanton
to $SU(4)$ rather than to a single $SU(2)$, in order
to be able to higgs the off-diagonal strings. By
an index theorem, there are $16$ real bosonic zero
modes. $8$ of them are W-bosons, and the
other $8$ are massive Higgs. The 8-8 strings
are discussed in \cite{Hor}.

We have identified the stretched strings between
a $p$-brane and a $p'$-brane as  just the zero
modes of off-diagonal blocks; one would like
to ask what about the massive modes of
$p$-$p'$ strings in the M(atrix) theory.
On one hand, for short open strings these
modes, similar to the massive modes
of short open 0-0 strings, are simply
absent in the M(atrix) model by postulate.
(It would be interesting to examine the
long strings in M(atrix) theory ending
on $p$ $p'$-branes.)  On the other hand,
it might be wise to leave the possibility
open that these massive modes on short strings
and other massive modes, such as KK modes in
a higher dimensional super Yang-Mills theory
could be physically relevant so that their
inclusion is necessary to make the high
dimensional theory well-defined in the UV regime.
We leave investigation of this issue to the future.

How about the higher modes of the off-diagonal
blocks? Could their effects approximate to those
of the massive modes of $p$-$p'$ strings? We do
not think so, since  the latter is graded by $\alpha'$,
while the former is determined by the scale of the
background field and the scale of the torus. The
modified M(atrix) model in the presence of
a longitudinal 5-brane proposed in Ref. \cite{BD}
should be viewed as a low-energy effective theory
of the fundamental M(atrix) model, in which the
higher modes of the off-diagonal block are ignored.
Indeed, in this case the zero-modes of the
off-diagonal block dominate the low-energy physics,
since surviving supersymmetry makes the
contributions of the higher modes cancel in the
leading order at large distances.

Although in this papers we have used
the \IIA language for brane names, the above
discussions are of M theory nature.
It may be amusing to consider an alternative \IIA theory
which is obtained by compactifying the ninth
direction and interchanging the role of the ninth
and eleventh directions. What we called D0-branes
above become short strings, which are also understood
as D0-branes by introducing unit electric flux to
the corresponding matrix element \cite{DVV}.
We leave the complete analysis and related topics
for the future.
%$D_0=\frac{\del}{\del t}+c\s$ and
%$D_9=\frac{\del}{\del\s}$ be the covariant
%derivatives giving the constant electric flux $c$.
%Then at the point $\s=0$ on the short string
%the zero modes we found above, which are
%independent of $t$ and $\s$, remain solutions
%of the equations of motion. It can also be
%checked that no new zero modes are introduced.
%Hence our results about zero-mode counting
%can be carried over to this \IIA theory.

\section{Acknowledgment}

The work of P.M.H. and Y.S.W. is supported in part by
U.S. NSF grant PHY-9601277. The work of M.L. is
supported by DOE grant DE-FG02-90ER-40560
and NSF grant PHY-9123780.


\vskip .8cm

\baselineskip 22pt

\begin{thebibliography}{10}

\itemsep 0pt

\bibitem{Pol}
J. Polchinski:
``Dirichlet-Branes and Ramond-Ramond Charges'',
hep-th/9510017;
{\em Phys. Rev. Lett.} {\bf 75}, 4724 (1995).

\bibitem{DLP}
J. Dai, R. G. Leigh, J. Polchinski:
``New Connections Between String Theories'',
{\em Mod. Phys. Lett.} {\bf A4}, 2073-2083
(1989).

\bibitem{LP}
H.  Lu and C. Pope,
``P-Brane Taxonomy",
hep-th/9702086.

\bibitem{Wit}
E. Witten:
``Bound States of Strings and p-Branes'',
{\em Nucl. Phys.} {\bf B460}, 335-350
(1996); hep-th/9510135.

\bibitem{HW}
P.M. Ho and Y.S. Wu:
``D-branes and Noncommutative Geometry",
{\em Phys. Lett.} {\bf B 398}, 251 (1997);
hep-th/9611233.

\bibitem{BFSS}
T. Banks, W. Fischler, S. H. Shenker, L. Susskind:
``M Theory as a Matrix Model: A Conjecture'',
{\em Phys. Rev.} {\bf D55}, 5112-5128 (1997);
hep-th/9610043.

\bibitem{BSS}
T. Banks, N. Seiberg, S. Shenker:
``Branes from Matrices'',
{\em Nucl. Phys.} {\bf B490}, 91-106 (1997);
hep-th/9612157.

\bibitem{GRT}
O. J. Ganor, S. Ramgoolam, W. Taylor IV:
``Branes, Fluxes and Duality in M(atrix)-Theory'',
hep-th/9611202.

\bibitem{BD}
M. Berkooz, M. R. Douglas:
``Five-Branes in M(atrix) Theory'',
{\em Phys. Lett.} {\bf B395}, 196-202 (1997);
hep-th/9610236.

\bibitem{DVVa}
R. Dijkgraaf, E. Verlinde, H. Verlinde:
``5D Black Holes and Matrix Strings'',
hep-th/9704018.

\bibitem{Pol1}
J. Polchinski:
``TASI Lectures on D-branes'',
hep-th/9611050.

\bibitem{KR}
N. Kim and S.J. Rey:
``M(atrix) Theory on an Orbifold and Twisted Membrane'',
hep-th/9701139.

\bibitem{BSS1}
T. Banks, N. Seiberg, E. Silverstein:
``Zero and One-Dimensional Probes with N=8 Supersymmetry'',
hep-th/9703052.

\bibitem{BM}
T. Banks, L. Motl:
``Heterotic Strings from Matrices'',
hep-th/9703218.

\bibitem{Hor}
P. Horava:
``Matrix Theory and Heterotic Strings on Tori'',
hep-th/9705055.

\bibitem{BR}
M. Berkooz, M. Rozali:
``String Dualities From Matrix Theory'',
hep-th/9705175.

\bibitem{Motl}
L. Motl:
``Proposals on Nonperturbative Superstring Interactions'',
hep-th/9701025.

\bibitem{BS}
T. Banks, N. Seiberg:
``Strings from Matrices'',
hep-th/9702187.

\bibitem{DVV}
R. Dijkgraaf, E. Verlinde, H. Verlinde:
``Matrix String Theory'',
hep-th/9703030.

\bibitem{Town}
P. K. Townsend:
``The Eleven-Dimensional Supermembrane Revisited'',
{\em Phys. Lett.} {\bf B350}, 184 (1995);
hep-th/9501068.

\bibitem{GR}
Z. Guralnik and S. Ramgoolam:
``Torons and D-Brane Bound States'',
hep-th/9702099.

\bibitem{tH}
G. 't Hooft:
``A Property of Electric and Magnetic Flux
in Non-Abelian Gauge Theories'',
{\em Nucl. Phys.} {\bf B153}, 141-160 (1979).

\bibitem{ASch}
A. S. Schwarz:
``On Regular Solutions of Euclidean Yang-Mills Equations'',
{\em Phys. Lett.}{\bf B67}, 172-174 (1977).\\
A. S. Schwarz:
``Instantons and Fermions in the Field of Instanton'',
{\em Comm. Math. Phys.}
{\bf 64}, 233-268 (1979).

\bibitem{Baal}
P. van Baal:
``$SU(N)$ Yang-Mills Solutions with Constant Field Strength on $T^4$'',
{\em Comm. Math. Phys.}
{\bf 94}, 397-419 (1984).

\bibitem{Zum}
B. Zumino:
``Euclidean Supersymmetry and the Many-Instanton Problem'',
{\em Phys. Lett.}{\bf B69}, 369-371 (1977).

\bibitem{AS}
M. F. Atiyah, I. M. Singer:
{\em Ann. Math.} {\bf 87}, 485 (1968);
{\em Ann. Math.} {\bf 87}, 546 (1968);
M. F. Atiyah, G. B. Segal:
{\em Ann. Math.} {\bf 87}, 531 (1968).\\
See, for example, M. Nakahara:
{\em Geometry, Topology and Physics},
Institute of Physics Publishing (1990)
for an introduction.

\bibitem{DF}
U. H. Danielsson, G. Ferretti:
``The Heterotic Life of the D-particle'',
hep-th/9610082.

\bibitem{Tay1}
W. Taylor IV:Adhering 0-Branes to 6-Branes and 8-Branes'',
hep-th/9705116.

\bibitem{ASJ}
H. Arfaei, M. M. Sheikh Jabbari:
``Different D-Brane Interactions",
hep-th/9608167.

\bibitem{LM}
M. Li and E. Martinec:
``Matrix Black Holes'',
hep-th/9703211;
``On the Entropy of Matrix Black Holes'',
hep-th/9704134.

\bibitem{JM}
J. Maldacena:
``Probing Near Extremal Black Holes with D-branes'',
hep-th/9705053.

\bibitem{Doug}
M. R. Douglas:
``Gauge Fields and D-Branes'',
hep-th/9604198.

\bibitem{Witten}
E. Witten: ``Sigma Models and the ADHM Construction of Instantons'',
{\em J. Geom. Phys.} {\bf 15}, 215-226 (1995);
hep-th/9410052.

\bibitem{Aharony}
O. Aharony and M. Berkooz: "Membrane Dynamics in M(atrix) Theory",
hep-th/9611215.

\bibitem{Lif1}
G. Lifschytz and S. Mathur: "Supersymmetry and
Membrane Interactions in M(atrix) Theory", hep-th/9612087.

\bibitem{Lif2}
G. Lifschytz: ``Four-brane and Six-brane Interactions
in M(atrix) Theory", hep-th/9612223

%\bibitem{Roz}
%M. Rozali:
%``Matrix Theory and U-duality in Seven Dimensions'',
%hep-th/9702136.

%\bibitem{Tay}
%W. Taylor IV:
%``D-Brane Field Theory on Compact Spaces'',
%hep-th/9611042.

\end{thebibliography}
\end{document}